\documentclass[twocolumn,showpacs,preprintnumbers,amsmath,amssymb,aps]{revtex4}
\usepackage{graphicx}
\usepackage{dcolumn}
\usepackage{bm}

\begin{document}

\title{Atomic fountain of laser-cooled Yb atoms for precision measurements}
 \author{Kanhaiya Pandey}
 \author{K. D. Rathod}
 \author{Alok K. Singh}
 \author{Vasant Natarajan}
 \email{vasant@physics.iisc.ernet.in}
 \homepage{www.physics.iisc.ernet.in/~vasant}
 \affiliation{Department of Physics, Indian Institute of
 Science, Bangalore 560\,012, INDIA}

\begin{abstract}
We demonstrate launching of laser-cooled Yb atoms in a cold
atomic fountain. Atoms in a collimated thermal beam are
first cooled and captured in a magneto-optic trap (MOT)
operating on the strongly-allowed ${^1S}_0 \rightarrow
{^1P}_1$ transition at 399~nm (blue line). They are then
transferred to a MOT on the weakly-allowed ${^1S}_0
\rightarrow {^3P}_1$ transition at 556~nm (green line).
Cold atoms from the green MOT are launched against gravity
at a velocity of around 2.5~m/s using a pair of green
beams. We trap more than $10^7$ atoms in the blue MOT and
transfer up to 70\% into the green MOT. The temperature for
the odd isotope, $^{171}$Yb, is $\sim$1~mK in the blue MOT,
and reduces by a factor of 40 in the green MOT.
\end{abstract}



\maketitle

\section{Introduction}
Laser cooling and trapping of Yb \cite{HTK99,RKW04,PRP10}
is different from the more common alkali-metal atoms
because of its spin-zero ground state. It has two cooling
transitions (one strong and one weak) with widely-differing
Doppler temperature limits, and does not require the use of
a repumping laser. In addition, it has 7 stable isotopes,
of which 5 are bosonic and 2 are fermionic. This allows the
comparative study of Fermi-Bose gas mixtures, particularly
under conditions of quantum degeneracy \cite{TMK03,FTK07}.
Furthermore, spin-exchange collisions in the closed-shell
ground state are smaller compared to the alkali-metal
atoms. This makes laser-cooled Yb an attractive candidate
for precision measurements and atomic clocks \cite{HBO05}.
One of us (VN) has recently proposed \cite{NAT05} using
laser-cooled Yb atoms launched in an atomic fountain for a
high-precision test of the existence of a permanent
electric dipole moment (EDM). The existence of an atomic
EDM would be direct evidence of time-reversal symmetry
violation in the laws of physics. Therefore, EDM searches
are among the most important atomic physics experiments as
they can strongly constrain theories that go beyond the
Standard Model.

\begin{figure}
\centering{\resizebox{0.9\columnwidth}{!}{\includegraphics{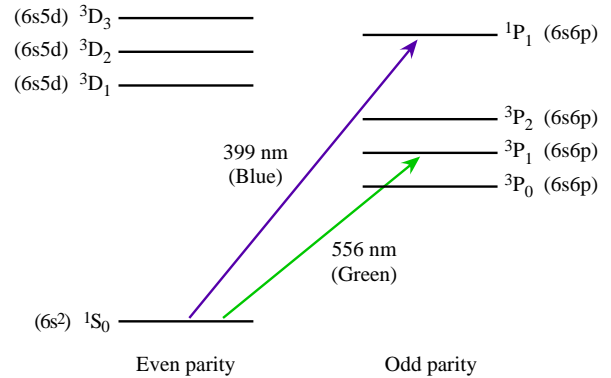}}}
\caption{(Color online) Low-lying energy levels of Yb
showing the two transitions that can be used for laser
cooling and trapping.}
 \label{levels}
\end{figure}

In this work, we demonstrate an atomic fountain of
laser-cooled Yb atoms for use in precision measurements.
The relevant low-lying energy levels of Yb are shown in
Fig.\ \ref{levels}. As in the case of alkaline-earth
elements, there are two transitions that can be used for
laser cooling. The strong one is the ${^1S}_0 \rightarrow
{^1P}_1$ transition at 399~nm (blue), and the weak one is
the ${^1S}_0 \rightarrow {^3P}_1$ intercombination line at
556~nm (green). The hot atoms emanating from an effusive
oven are first slowed using a Zeeman slower and then
captured in a magneto-optic trap (MOT), both operating on
the strong transition. This transition is similar to the
laser-cooling transitions in the alkali-metal atoms and has
a large capture velocity. We can trap more than $10^7$ Yb
atoms in the blue MOT, at a temperature of $\sim$3~mK for
the even isotopes and $\sim$1~mK for the odd isotopes. The
weak transition has a $150 \times$ smaller linewidth, which
gives a lower MOT temperature but also means the capture
velocity is quite small. This makes it difficult to
directly load the green MOT from the Zeeman-slowed beam. We
therefore transfer atoms captured in the blue MOT into the
green MOT. Using a multi-step transfer process, we transfer
up to 70\% of the atoms and obtain a temperature that is 40
times smaller. The cold atoms from the green MOT are
launched against gravity using a pair of green beams in
{\it moving-molasses} configuration \cite{CSG91}. The
launch velocity is varied from 2.1 to 2.9~m/s by adjusting
the detuning of the moving-molasses beams.

\section{Experimental details}
\begin{figure}
\centering{\resizebox{0.99\columnwidth}{!}{\includegraphics{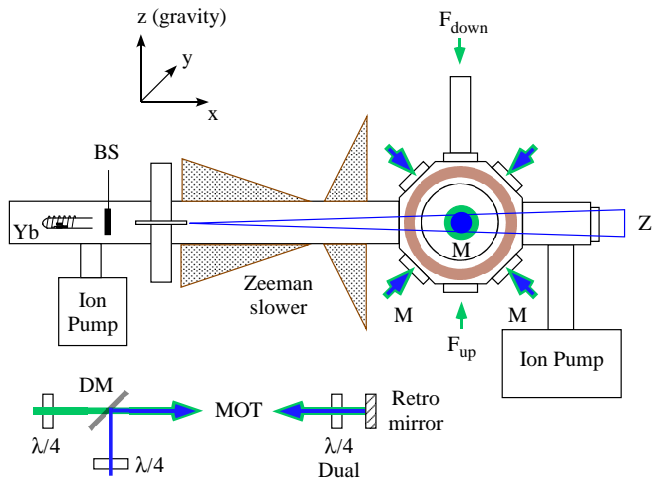}}}
\caption{(Color online) Experimental chamber. Gravity is
along $z$, and the MOT axis is along $y$. The 3 sets of MOT
beams consist of both blue and green beams, and are
produced using a dichroic mirror (DM) as shown in the
bottom. Figure abbreviations: BS -- beam shutter, M --
incoming MOT beam, F$_{\rm up,down}$ -- fountain beams, Z
-- Zeeman slowing beam, $\lambda/4$ -- quarter waveplate.}
 \label{chamber}
\end{figure}

The main experimental chamber, shown schematically in Fig.\
\ref{chamber}, consists of three regions: a source region,
a Zeeman slower region, and the MOT/fountain region. The
source is a quartz ampoule that is resistively heated to
about 400$^\circ$C. The ampoule contains elemental Yb with
all the isotopes in their natural abundances. The atomic
beam is collimated using a copper skimmer with a small
hole, and has a pneumatic shutter in front of it. The
source region is pumped with a 20~l/s ion pump so that the
pressure is below $10^{-7}$~torr when the source is on.

The source region is connected to the Zeeman-slowing region
using a small differential pumping tube, which allows the
experimental chamber to be at much lower pressure and also
provides further collimation. The slower is a tube of 40~mm
diameter by 500~mm length. The required magnetic field
profile is generated by winding welding cables on the
outside. At the end of the slower is the main MOT chamber.
The distance to the center of the MOT is 220~mm. The MOT
chamber consists of 8 small (70-mm diameter) ports in the
$x-z$ plane and 2 large (100-mm diameter) ports along the
$y$ direction. The ports along $z$, which is the direction
of gravity, have the fountain chamber. The ports along $x$
are used to connect to the Zeeman slower on one side and a
55~l/s ion pump on the other side. The pressure in the MOT
region is below $10^{-9}$ torr. Optical access is provided
by glass viewports, while the fountain chamber is a
rectangular glass cell.

From the energy-level diagram, we see that the experiment
requires two main lasers. The blue beam at 399~nm is
produced by using a ring Ti:sapphire laser (Coherent
899-21) operating at 798~nm and doubling its output in an
external delta-cavity doubler (Laser Analytical Systems
LAS100). The output power of the Ti:sapphire is 1.5~W, and
its frequency is stabilized on a reference cavity to give
an rms linewidth of 1~MHz. The output of the doubler is
about 180 mW. Of this, 30--40~mW is sent through an
acousto-optic modulator (AOM) for the Zeeman-slowing beam.
The remaining power is used to produce the MOT beams. A
second low-power blue beam used for probing is generated
using a 30~mW diode laser (Nichia Corporation) stabilized
in an external grating-feedback cavity. The linewidth after
stabilization is about 1~MHz and the output power is 9~mW.
The green beam at 556~nm is produced by doubling the output
of a fiber laser operating at 1111~nm (Koheras Boostik
Y10). The output power of the fiber laser is 500~mW with a
linewidth of 70~kHz. The frequency is doubled in an
external ring-cavity (Toptica Photonics) with a
temperature-tuned KNbO$_3$ nonlinear crystal. The output
power of the doubler is 65~mW.

The MOT is made of three sets of counter-propagating beams,
two in the $x-z$ plane and one along the $y$ direction. The
quadrupole magnetic field is along the $y$ direction. Each
MOT beam is composed of a circularly-polarized blue beam
and a circularly-polarized green beam, mixed on a dichroic
mirror. The combined beams retroreflect on the other side
of the chamber through a dual-wavelength $\lambda/4$
waveplate.

The frequency of the blue laser from the doubler is
manually adjusted to maximize the MOT fluorescence and then
left untouched during the MOT loading time of a few
seconds. Since the drift of the Ti:sapphire laser is less
than 10~MHz/hour, the laser does not need to be actively
locked. On the other hand, the blue diode-laser beam used
for probing and the green beam are locked to their
respective transitions. The spectroscopy for locking is
done in a separate vacuum chamber that also has a
collimated Yb atomic beam. The laser beams are sent
perpendicular to the atoms, and the fluorescence is
collected by two photomultiplier tubes (PMT, Hamamatsu
R928). The blue laser is locked using modulation-free
locking \cite{YBD00}. The error signal is generated by
taking the difference between opposite circular
polarizations in the presence of a magnetic field (circular
dichroism). The green laser is locked to a peak by
frequency modulation at 20~kHz and lock-in detection to
generate the error signal.

\section{Results and discussion}

\subsection{Blue MOT}
We first discuss the loading of the blue MOT. The hot atoms
emanating from the source have a longitudinal velocity
distribution with a most-probable velocity of 310~m/s. All
the atoms with velocity below 250~m/s (which represents
about 30\% of the total number) are slowed down using a
spin-flip Zeeman slower \cite{CHI02}. The slower consists
of a decreasing field part near the beginning and then an
increasing field region near the end. The total slowing
distance is 450~mm, and the magnetic field varies from
210~G at the beginning to $-235$~G at the end. The slower
beam is focussed with a lens so that it has a size of 20~mm
near the MOT and 4~mm at the differential pumping tube. The
total power in the slowing beam is 20~mW and it is detuned
by $-330$~MHz from resonance.

We define a capture velocity $v_c$ such that atoms having
velocity below $v_c$ are cooled and captured in the MOT.
From a simple one-dimensional laser-cooling model, $v_c$ is
the velocity at which the Doppler shift takes the atom out
of resonance by one linewidth, therefore it is given by
\cite{RWN01},
\begin{equation}
v_c = \left( |\Delta| + \Gamma \right) \frac{\lambda}{2\pi} \, ,
 \label{vc}
\end{equation}
where $\Delta$ is the detuning of the beams. The value of
$\Gamma$ is $2\pi \times 28$~MHz for the blue transition,
hence $v_c$ is 22~m/s for a typical detuning of $\Gamma$.
Therefore, the Zeeman slower is designed to have a final
velocity of 20~m/s, so that all the slowed atoms are loaded
into the blue MOT.

The six MOT beams have a total power of 120~mW and size of
15~mm each. The detuning is optimized by looking at the MOT
fluorescence, and is around $-40$~MHz at an axial field
gradient of 30~G/cm. The ${^1S}_0 \rightarrow {^1P}_1$
transition is not closed since atoms can be lost to the
metastable ${^3P}_{0,2}$ states through the intermediate
$D$ states (see Fig.\ \ref{levels}). There are also losses
due to background collisions in the vacuum chamber. These
loss mechanisms limit the trap loading time constant to
1~s. We therefore load the trap for a total time of 2~s.
This gives a cold cloud of size 3~mm. By calibrating the
MOT fluorescence measured by a photodiode, we estimate the
number of atoms to be $10^7$.

\begin{figure}
\centering{\resizebox{0.95\columnwidth}{!}{\includegraphics{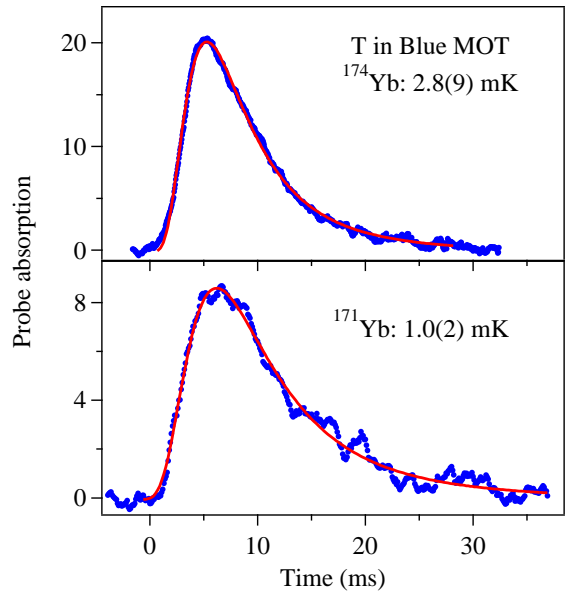}}}
\caption{(Color online) Absorption profile of a blue probe
beam used to measure the temperature in the blue MOT. The
solid curve is a fit to Eq.\ \ref{motT} with the
temperature as the only fit parameter.}
 \label{motb}
\end{figure}

The temperature in the MOT is measured by mapping the
velocity distribution, which in turn is determined by the
absorption of a probe beam using time-of-flight. The probe
beam is placed 6~mm below the trap center. If the
temperature of the MOT is $T$ and the distance to the probe
beam is $d$, then the absorption as a function of time $t$
is given by \cite{MAR03}
\begin{equation}
A(t) \propto \frac{d^3}{t^4} \exp\left( -\frac{M(d/t)^2}{2k_BT} \right) \, ,
 \label{motT}
\end{equation}
where $M$ is the mass of the atom, and $k_B$ is the
Boltzmann constant. In Fig.\ \ref{motb}, we show the
absorption profile of atoms after they are released from
the MOT. The solid curve is a fit to Eq.\ \ref{motT} with
the temperature as the only fit parameter. For the even
isotope $^{174}$Yb with zero nuclear spin, the temperature
in the blue MOT is 2.8(9)~mK. This is reasonable because
the Doppler limit is 0.6~mK, and it is known that the MOT
temperature is typically a few times higher than this limit
due to additional heating mechanisms \cite{CMK05}. For the
odd isotope $^{171}$Yb with $I=1/2$, the presence of
magnetic sublevels allows for sub-Doppler cooling. As a
result, the measured temperature of 1.0(2)~mK is a factor
of 3 lower.

\subsection{Green MOT}
One would think that the green MOT can be easily loaded
from the Zeeman-slowed beam. While this has been done
before \cite{KHT99}, direct loading is complicated by the
narrow linewidth of 182~kHz for this transition. Hence the
capture velocity (from Eq.\ \ref{vc}) is less than 1~m/s,
even if the detuning is $9 \, \Gamma$. If the atoms coming
out of the slower have this small velocity, the end of the
slower has to be less than 10~cm from the MOT center so
that atoms do not fall out of the trapping region under the
influence of gravity. We have in fact designed such a
chamber and loaded atoms into the green MOT directly from
the slowed beam. But the number of atoms in the MOT is much
less than what we get by first capturing atoms in the blue
MOT, and then transferring them into the green MOT. In the
following, we discuss this transfer method in detail.

\begin{figure}
\centering{\resizebox{0.95\columnwidth}{!}{\includegraphics{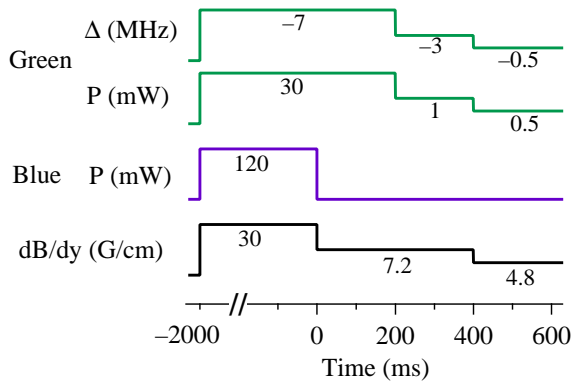}}}
\caption{(Color online) Sequence for transferring atoms
from the blue MOT to the green MOT.}
 \label{seqG}
\end{figure}

One important difference between the two MOTs is that the
optimal field gradient for the green MOT is much smaller
than that for the blue MOT. In addition, the transfer
efficiency and final MOT temperature are dependent on the
detuning and power of the green beams. Therefore the
transfer is done in a multi-step process as shown in the
sequence in Fig.\ \ref{seqG}. After the blue MOT is loaded
for 2~s, the blue beams are turned off at $t=0$.
Simultaneously the field gradient is lowered from 30 to
7~G/cm. At this time, the green MOT beams have a total
power of 30~mW and detuning of $-7$~MHz. After 200~ms, the
total power is lowered to 1~mW and the detuning to
$-3$~MHz. Finally, after another 200~ms, the total power
and detuning are set to their final values of 0.5~mW and
$-0.5$~MHz. The percentage transfer is measured by turning
the blue MOT beams back on. The fluorescence level at the
turn-on point is 0 without the green MOT, and jumps to some
fraction of its original value with the MOT on, indicating
that these many atoms have survived in the green MOT. Under
optimal conditions, we can transfer 70\% of atoms from the
blue to the green MOT.

\begin{figure}
\centering{\resizebox{0.95\columnwidth}{!}{\includegraphics{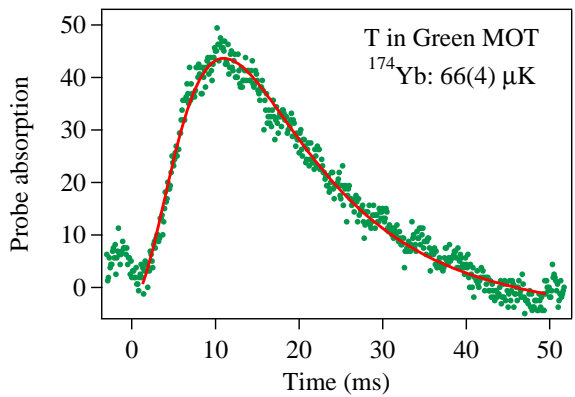}}}
\caption{(Color online) Absorption of a blue probe beam to
measure temperature in the green MOT. The solid curve is a
fit to Eq.\ \ref{motT} with the temperature as the only fit
parameter.}
 \label{motg}
\end{figure}

The main advantage of the green MOT is the lower final
temperature attained by the atoms. This is because the
Doppler limit for this narrow transition is only
4.4~$\mu$K. This advantage is evident from the temperature
measurement shown in Fig.\ \ref{motg}. For the even isotope
$^{174}$Yb with no sub-Doppler cooling, the temperature is
66(4)~$\mu$K, a factor of 40 lower than that in the blue
MOT. The temperature for the odd isotopes is not shown, but
is again 3 times smaller due to sub-Doppler cooling
\cite{MWR03}.

\subsection{Atomic fountain}
The final experiment was to launch the atoms from the green
MOT in an atomic fountain. The sequence for launching is
the same as that shown in Fig.\ \ref{seqG} up to the point
of loading of the green MOT, i.e., first loading of the
blue MOT for 2~s, then turning off the blue beams and
turning down the magnetic-field gradient at $t=0$, and then
lowering the power and detuning of the green MOT beams in
two steps to get a low final temperature. The launching is
done at $t=700$~ms by turning off all the beams and pulsing
on the green launching beams for 10~ms. The launched atoms
are probed by monitoring the absorption of a blue probe
beam placed 19~cm above the MOT center.

\begin{figure}
\centering{\resizebox{0.95\columnwidth}{!}{\includegraphics{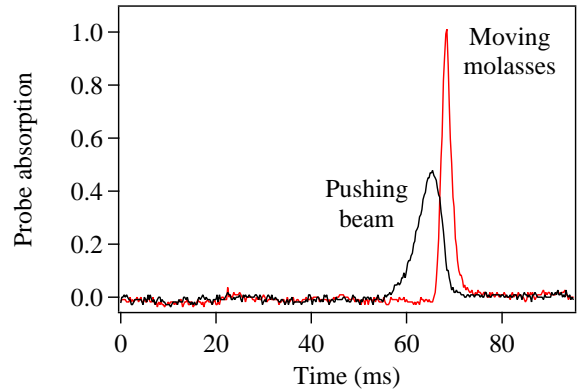}}}
\caption{(Color online) Absorption of a probe beam for
atoms launched either with a single pushing beam or two
beams in moving-molasses configuration.}
 \label{fount1}
\end{figure}

The launching can be done with just a single pushing beam
or using the idea of moving molasses \cite{CSG91}. With
just a pushing beam pointing up, the atoms get heated along
this direction. Instead, by using another beam pointing
down, the detunings can be chosen such that the atoms are
cooled in the launch direction. If we want the detuning to
be $-\Gamma/2$ (which gives the lowest temperature in
one-dimensional molasses) in a frame moving up with a
velocity $v$, then the detunings in the laboratory frame
are
\begin{equation}
\Delta_{\rm up} = \Gamma/2 - v/\lambda \ \ \ \
{\rm and \ \ \ \ }
\Delta_{\rm down} = \Gamma/2 + v/\lambda \, .
 \label{moving}
\end{equation}
The difference between the two methods of launching is seen
in Fig.\ \ref{fount1}. The absorption of the probe beam as
a function of time gives a measure of the longitudinal
velocity distribution. There is both a narrowing and an
increase in amplitude with the use of moving molasses,
clearly indicating cooling in this direction.

The low capture velocity of the green transition, which
complicates the direct loading of the MOT, is also a
problem with the launching. For a launch velocity of
$v=2.5$~m/s, $v/\lambda$ in Eq.\ \ref{moving} is 4.5~MHz,
which is 25 times the natural linewidth. Therefore, atoms
with 0 velocity will be outside the ``capture range'' of
the beams. To solve this problem, we ramp the detuning of
the launch beams from 0 to their final values using AOMs
over the launch period of 10~ms. This ensures that the
atoms are accelerated adiabatically from 0 to $v$. The
power in each beam is 1~mW, and they are brought to the
apparatus using single-mode fibers so that there is no
change in the direction as the detuning is ramped.

\begin{figure}
\centering{\resizebox{0.95\columnwidth}{!}{\includegraphics{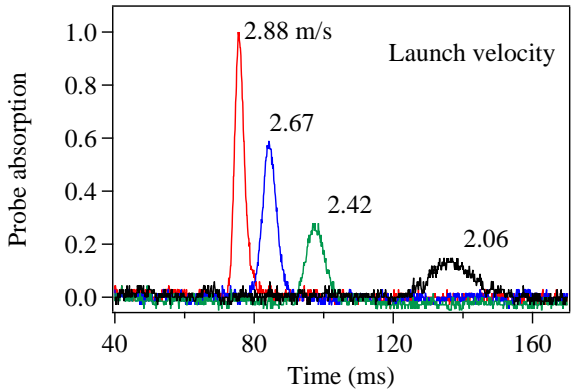}}}
\caption{(Color online) Absorption of a probe beam for
$^{171}$Yb atoms launched with different velocities.}
 \label{fount2}
\end{figure}

As seen from Eq.\ \ref{moving}, the launch velocity can be
changed by adjusting the final detunings of the up and down
launch beams. In Fig.\ \ref{fount2}, we show the results of
launching $^{171}$Yb atoms at four velocities, ranging from
2.06~m/s to 2.88~m/s. A change in relative detuning
$\Delta_{\rm up} -  \Delta_{\rm down}$ of 800~kHz
corresponds to a change in launch velocity by 0.22~m/s. As
the velocity decreases, the transverse and longitudinal
temperatures become more important since the time to reach
the probe beam is longer. As a result, there is increased
spread in the signal.

\section{Conclusion}
In conclusion, we have demonstrated launching of
laser-cooled Yb atoms in an atomic fountain. The hot atoms
emanating from a thermal source are Zeeman slowed and
captured in a MOT on the strongly-allowed blue transition.
Laser cooling on this transition is similar to cooling of
the more common alkali-metal atoms. We capture more than
$10^7$ atoms in the blue MOT, with a temperature of about
3~mK for the even isotopes and 1~mK for the odd isotopes.
The other laser cooling transition in Yb is a
weakly-allowed green transition, which provides not only
unique opportunities but also its own problems. In
particular, direct loading of the green MOT from a
Zeeman-slowed beam is complicated by its small capture
velocity. We therefore first capture atoms in the blue MOT
and then transfer to the green MOT. The green transition
gives a factor of 40 lower temperature in the MOT, but to
achieve this the transfer has to be done in a multi-step
process with progressively smaller detuning and power in
the trapping beams. Under optimal conditions, we can
transfer 70\% of the atoms into the green MOT.

The ultracold atoms from the green MOT are launched in an
atomic fountain using a pair of green beams in the vertical
direction in {\it moving-molasses} configuration. This
allows us to control the launch velocity without additional
heating in the launch direction. Launching atoms in a
fountain is again more difficult compared to the
alkali-metal atoms because of the small capture velocity of
the green transition. We overcome this problem by
adiabatically ramping the detuning in the moving-molasses
beams, and are then able to launch atoms with velocities
varying from 2.1~m/s to 2.9~m/s. The cold atoms can be used
for precision measurements such as optical clocks and the
search for a permanent atomic EDM. We have recently used a
thermal beam of Yb atoms for spectroscopy using the Ramsey
separated-oscillatory-fields technique. The application of
this idea to a cold atomic fountain should improve the
sensitivity by several orders of magnitude.

\begin{acknowledgments}
This work was supported by the Department of Science and
Technology, India. K.P., K.D.R., and A.K.S. acknowledge
support from the Council of Scientific and Industrial
Research, India.
\end{acknowledgments}


\end{document}